\documentclass[10pt,letter]{article}
\usepackage[utf8]{inputenc}
\usepackage{geometry}
\usepackage{times} 

\usepackage{moreverb,url}
\usepackage{authblk}
\usepackage{ulem}
\usepackage[colorlinks,bookmarksopen,bookmarksnumbered,citecolor=red,urlcolor=black]{hyperref}
\usepackage{epsfig}

\title{Oscillating Middle Vehicle in a Platoon}

\author[1]{Aditya Ghawre}
\author[2]{Eric Jacuzzi}
\author[1]{Kenneth Granlund}
\affil[1]{North Carolina State University, Raleigh, 27695, NC, USA}
\affil[2]{NASCAR R\&D, Concord, 28027, NC, USA}

\date{}

\begin{document}
\maketitle

\section{Introduction}
Platooning is an emerging concept where vehicles are driven at close proximity to each other, maintaining that distance throughout the journey. Several studies were carried out to determine how drag is varied and how it can be reduced using this concept. However, drag reduction is not a given in any platoon configuration. Reduction in the drag of a platoon depends on several factors, such as position of the vehicles, separation distance, platoon speed, incoming wind angle, number of vehicles, shape of the vehicles, etc. Platooning is especially effective for groups of bluff bodies such as cycling pelotons \cite{blocken2018}. Single vehicle drag optimization is a common practice for nearly all commercial vehicles. However, it is often done in isolation, neglecting interactions with other vehicles in a highway environment.
Various studies have been carried out to assess the feasibility, effectiveness, and challenges of platooning using a variety of vehicle models. Among the shapes that are heavily researched, the Ahmed body \cite{ahmed1984} is a simplified body resembling the shape of a van. Most studies of the Ahmed body or platooning have a heavy emphasis on the wake turbulence generated behind the body. In reality, turbulence is an integral part of nature and it significantly affects the drag on vehicles depending on their geometry \cite{watkins2007,gheyssens2016}. Because of its simplicity in design and readily available data, this study utilizes three Ahmed bodies to form a platoon. It has been shown that for an Ahmed body, drag decreases with increasing backlight angle until reaching the critical backlight angle ($30^\circ$) \cite{vino2005,pagliarella2007}. Longitudinal vortices formed at the rear of the Ahmed body have different intensities depending on the backlight angle and tend to move outwards as the separation distance increases \cite{pagliarella2006}. When platoons with varying numbers of vehicles were compared, it was found that increasing the number of platoon members decreased drag \cite{zabat1994,zabat1995,mcauliffe2018}. For the trailing vehicle in the platoon, spacing becomes critical because of the occurrence of resonance at 0.1 to 0.5 car length spacing and needs to be avoided as it increases the drag significantly for the trailing vehicle. Additionally, having a stationary ground during simulation can lead to erratic results because of the generation of the ground boundary layer. This can be ignored if the ground clearance of the model is greater than 10 times the boundary layer thickness \cite{zabat1994}. This study \cite{zabat1995} also showed that results of a four-car platoon can be extrapolated for larger platoons. Along with the drag reduction, lift is also affected by platooning \cite{romberg1971}. Some studies compared the drag benefits of different platoons having vehicles with different geometries \cite{legood2018,siemon2018}. Drag of streamlined platoons was greater than that of a streamlined vehicle traveling alone. This further enforces that vehicles which are designed to perform effectively in isolated conditions are more likely to cause drag penalties in the platoon. It also showed that the geometry of vehicles is an important factor only for spacing less than one car-length. For greater spacing, drag savings of the platoon does not depend on the geometry. Schito \cite{schito2012} concluded that cars having sharp back angles experience lower drag than fastback cars in a platoon. Reducing the spacing from 2 to 0.5 car-lengths reduces the drag. In a platoon, drag remains almost the same for all the members after the fourth, except for the last vehicle which shows an increase due to lower wake pressure.
In a study performed on the truck platooning, it was found that driving a tractor-trailer combination in a platoon can achieve the same drag benefits as that of a single aerodynamic tractor-trailer combination \cite{ellis2015}. Salari \cite{salari2018} showed that engine cooling air supply is reduced for lower spacing. Misalignment of the trailing vehicles with up to 50\% car-width does not affect the drag significantly. In another study \cite{vegendla2015}, which focused on yaw averaged aerodynamic drag (YAD) which is the average drag of $0^\circ$, $-6^\circ$, and $6^\circ$ yawed conditions, it was concluded that YAD reduction decreases with increasing the separation distance. In the case of side-by-side platoons, leading vehicles experience increased YAD. On-road testing of two trucks showed an improvement in fuel economy with decreasing the separation distance \cite{humphreys2016a}.  Lateral offset of the leading vehicle reduces the platoon performance at closer spacing.  For very close spacing, the rear truck showed penalty in drag \cite{humphreys2016b,hong1998}. From a study being carried out under the project ‘Safe Road Trains for the Environment (SATRE),’ it was observed that a four-vehicle platoon gives an average of 20\% drag reduction \cite{davila2011}. Fuel savings are increased with decreased separation distance \cite{davila2013a,davila2013b}. In a three-car platoon, the middle vehicle shows the greatest fuel savings whereas the lead vehicle might consume more fuel \cite{michaelian2001}. Reduction in pollution is achieved by drag reduction which is enhanced by increasing the number of vehicles in a platoon \cite{mitra2007}. Platooning, or drafting is significantly seen in motorsports, especially in the National Association for Stock Car Auto Racing (NASCAR) where cars have very similar drag and engine power values. A study on drafting observed in NASCAR race vehicles was conducted by Jacuzzi \cite{jacuzzi2019} to modify the drag performance of the vehicles during drafting situations. Certain trailing vehicle positions were observed to have increased drag compared to isolated vehicle drag, particularly in the 0.5-1.0 car length spacing. It was found out that using passive ducting from the vehicle nose out of the front wheel opening to widen the wake significantly reduces the maximum drag experienced by the trailing car at a spacing of one car length.
These results make the basis of this study that steady-state analysis of platooning behavior may not give the most accurate real-time results. Oscillations can cause phase shifting phenomenon resulting in different outcomes, e.g. increased/decreased maximum/minimum forces, etc. Therefore, a need for studying the transient effects of automotive aerodynamics arises. When a model in a platoon of vehicles starts oscillating longitudinally, it faces different incoming velocity than the non-oscillating members. As the drag varies with the square of this velocity, it experiences a different drag coefficient than the rest of the members. It is also interesting to see the phase shifting for such oscillations. Especially when the vehicles are driven very close to each other (~0.1 to 0.3L, where L is the length of the vehicle), drag drastically increases \cite{zabat1994}. In this case, if a member is oscillating, the time-averaged drag may not necessarily agree with a time-instantaneous drag due to flow separation convective effects. Since bluff bodies are very much vulnerable to drag at higher speeds, it was decided to conduct investigations on the bluff geometry of the Ahmed body as a methodology test for whether time-dependent simulations of moving bodies are required.

\section{Research cases}
\subsection{Single vehicle in steady-state conditions}
As a base case, CFD simulations were carried out for a single standard Ahmed body at the Reynolds number of $2.86 \cdot 10^5$ based on the length L of the body. The effects of platooning under different conditions, as mentioned below, are compared against this case. In this study, this case is referred to as the steady standalone case.
\subsection{Platoon vehicle in steady-state conditions}
\begin{figure}
\centering
\includegraphics[width=\columnwidth]{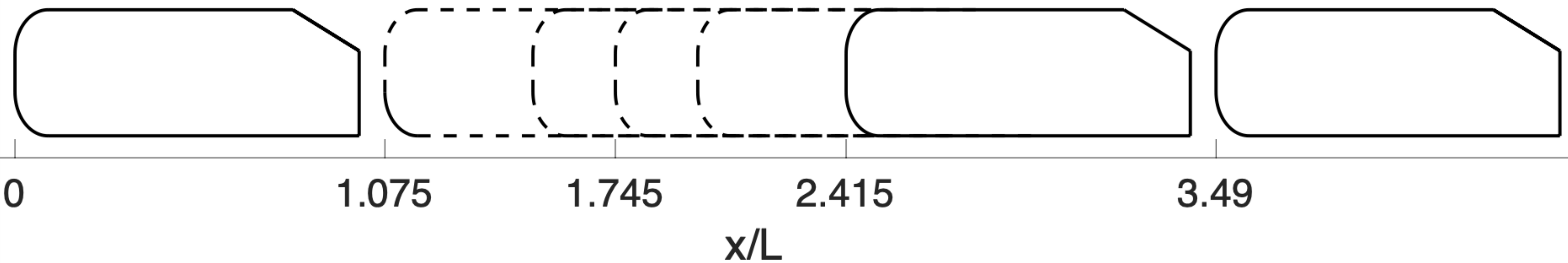}
\caption{Platoon with steady cases. The middle vehicle is shown in five different positions with Case 1 most rearward at x/L=2.415}
\label{figure1}
\end{figure}

A platoon of three vehicles is studied where the middle vehicle was varied over five different static positions. The position of the front of the leading vehicle is at x/L=0, and the tail is at x/L=1; the front of the middle vehicle mean position is at x/L=1.745 and the tail is subsequently at x/L=2.75; the front of the trailing vehicle is at x/L=3.49 and the tail is at x/L=4.49, as shown in Figure \ref{figure1}. These results were compared with the transient drag results to check the effect of transient nature of the aerodynamic drag.
Case 1 has the middle member at the rearmost position. The gap between the front end of the trailing member and the rear end of the middle member is the closest, 0.075L. Case 2 has the middle member at an intermediate location between its mean and the rearmost position. The gap between the rear of the middle member and the front of the trailing member is increased to 0.50L. In Case 3, the middle member is maintained at its mean position. The separation gaps between the neighboring vehicles are kept at 0.75L. Case 4 and 5 are parts of the forward cycle, i.e. the middle member is moving forward. It is at an intermediate position between its mean and the foremost position where the gap between the leading and the middle member is 0.50L. Case 5 shows the middle member at the foremost position with the closest gap between the front and the middle member being 0.075L.
\subsection{Platoon vehicles in transient conditions}
Two transient cases are considered which correlate to NASCAR drafting scenarios where a middle vehicle in a platoon oscillates in position. We can define reduced frequency k from the frequency of oscillation f, the vehicle length L and the velocity $U_\infty$
\begin{equation}
    k=\frac{\pi f L}{U_\infty}
\end{equation}
Here we investigate two dynamic cases, both with a vehicle length L of 5m, oscillated at a frequency f of 1Hz at an amplitude A/L of 0.67 (from Figure \ref{figure1}) at two different velocities $U_\infty$ of 78 m/s (175 mph) and 38.8 m/s (87 mph), resulting in the reduced frequencies of k=0.2 and k=0.4, respectively. At full scale, this results in a Reynolds number of $26 \cdot 10^6$ and $13 \cdot 10^6$ respectively, for which bluff body drag results are Reynolds number independent. In order to ensure Reynolds number independency at the computational scale, both oscillation cases are computed at the same aforementioned Reynolds number of $2.86 \cdot 10^5$, but varying oscillation frequencies in order to match the reduced frequencies. The computational investigations are therefore conducted with oscillation frequencies of f=0.25 Hz and f=0.5 Hz respectively.

\section{Computational Analysis}
\begin{figure}
\centering
\includegraphics[width=\columnwidth]{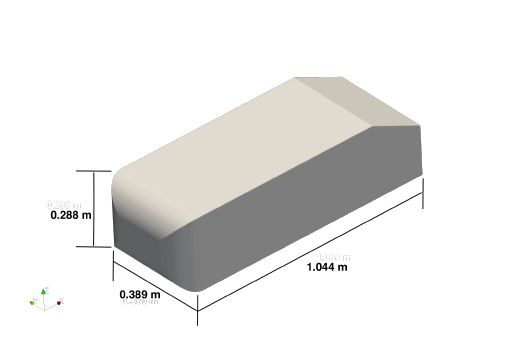}
\caption{$25^\circ$ slant back Ahmed body}
\label{figure2}
\end{figure}

\begin{figure}
\centering
\includegraphics[width=\columnwidth]{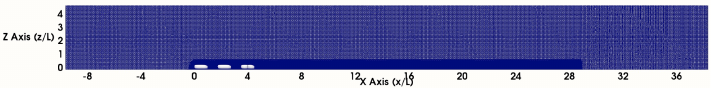}
\caption{Side view of the section of the decompressed grid normalized with vehicle length L}
\label{figure3}
\end{figure}

\begin{figure}
\centering
\includegraphics[width=\columnwidth]{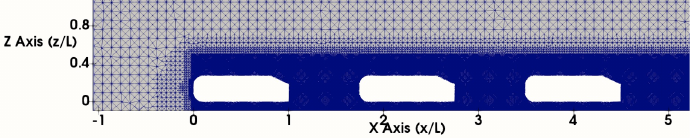}
\caption{Wake refinement and transition region shown with the zoomed-in decompressed grid normalized with vehicle length L}
\label{figure4}
\end{figure}
Computational simulations are performed on TotalSim’s version of OpenFOAM, using the full-scale Ahmed body with a 25° back glass slant, shown in Figure \ref{figure2}. The Ahmed body \cite{ahmed1984} dimensions are 1.044 x 0.389 x 0.288 meter (length x width x height). The computational domain is size is normalized with the Ahmed body’s length, given as 95.78L x 19.15L x 19.25L. The cross-sectional area of the Ahmed body is $0.1120 m^2$, equating to a domain blockage ratio of 0.028\%. The road surface is modeled as a moving boundary with velocity equal to the free stream velocity of 4.056 m/s. The resulting Reynolds number is $2.86 \cdot 10^5$ using the Ahmed body length L as the reference. Front wall is a velocity inlet with freestream velocity imposed and a turbulence intensity of 0.2\%. The grid is an unstructured hexahedral volume with a mix of tetrahedral and polyhedral elements at transition regions between the prism layers and volume size refinement interfaces. Surface mesh is modeled at 0.5 mm resulting in $y^+<1$, with a consistent wake refinement region as shown in Figures \ref{figure3} and \ref{figure4}. Typical mesh size is on the order of 100 million, consistent with the domain used by Jacuzzi and Granlund \cite{jacuzzi2019}.
Simulations consisted of an initial steady state, Reynolds Averaged Navier-Stokes (RANS) simulation followed by a transient simulation commencing from the steady state data. Simulations use an incompressible density formulation and Shear-Stress Transport (SST) $k-\omega$ turbulence model. Custom wall functions developed by TotalSim USA are used to resolve the wall within the three-cell prism layer. The wall function method by Kalitzin \cite{kalitzin2005} has been implemented by Ludlow \cite{ludlow2009} to improve turbulence modeling accuracy in automotive applications. During the initial phase of Detached Eddy Simulation (DES), the middle vehicle is stationary for t’=1.94 to establish a baseline, where t’ is the convective time normalized by the following equation:
\begin{equation}
    t^\prime=\frac{t U_\infty}{L}
\end{equation}
After the steady-state portion of the simulation a Detached Eddy Simulation (DES) computation of $t^\prime = 11.65$ commences, with a time step of 0.00038 convective units. Middle body motion is defined via a sinusoidal variation in position that avoids instantaneous acceleration of the middle body. During movement of the middle body, the intervening grid is compressed with the movement of the middle vehicle. Care is taken using this approach to ensure the boundary layer grids are not influenced by the compression of cells between bodies, and that the intervening cells do not become overly skewed. Maximum skewness was calculated to be 0.68 which occurs at the maximum compression time-step.

\section{Results}
\subsection{Single vehicle vs steady vehicles in a platoon}
\begin{figure}
\centering
\includegraphics[width=\columnwidth]{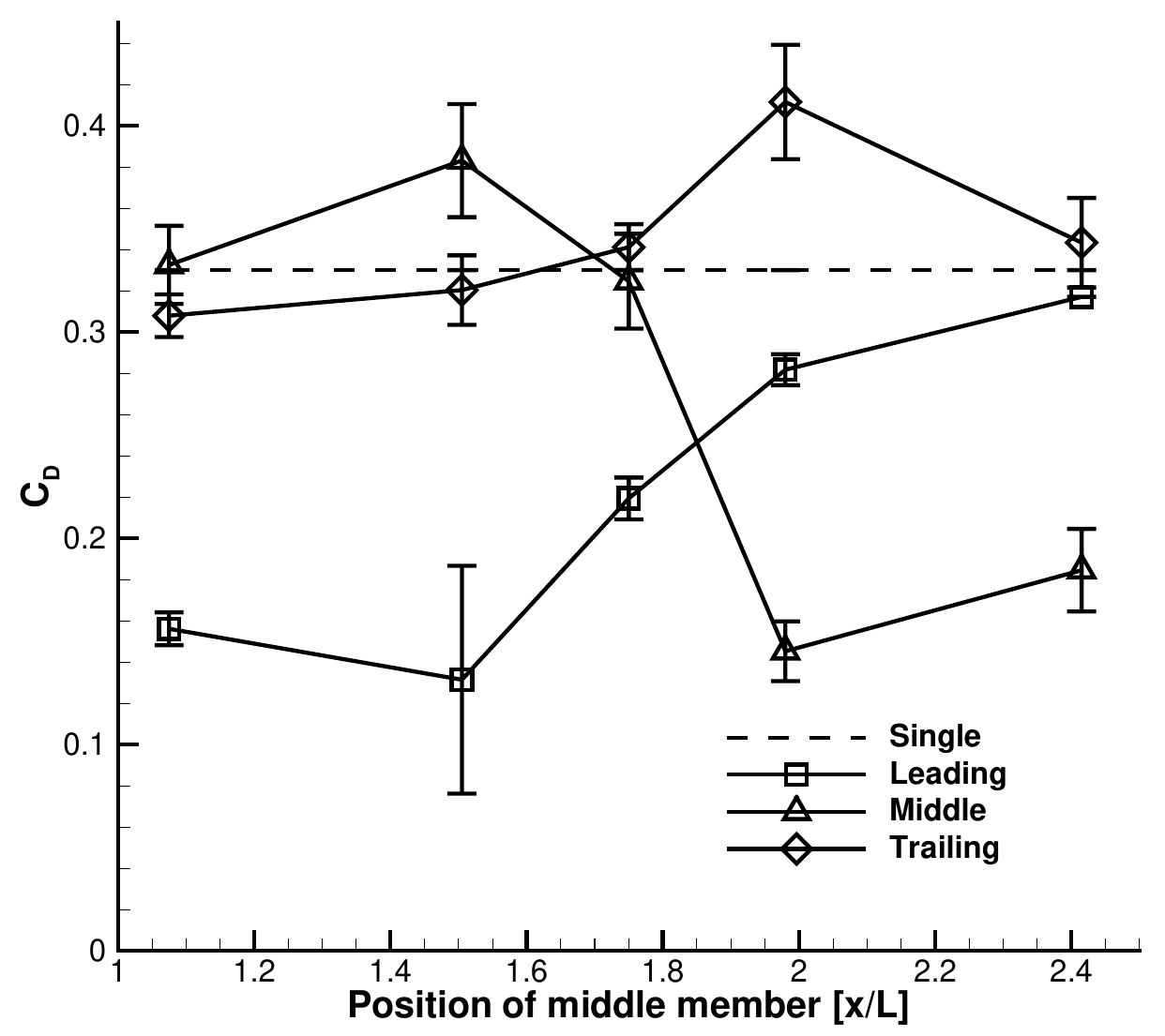}
\caption{Comparison of coefficient of drag of single vehicle with the steady cases. Errorbars indicate standard deviation of fluctuation.}
\label{figure5}
\end{figure}
\begin{figure}
\centering
\includegraphics[width=\columnwidth]{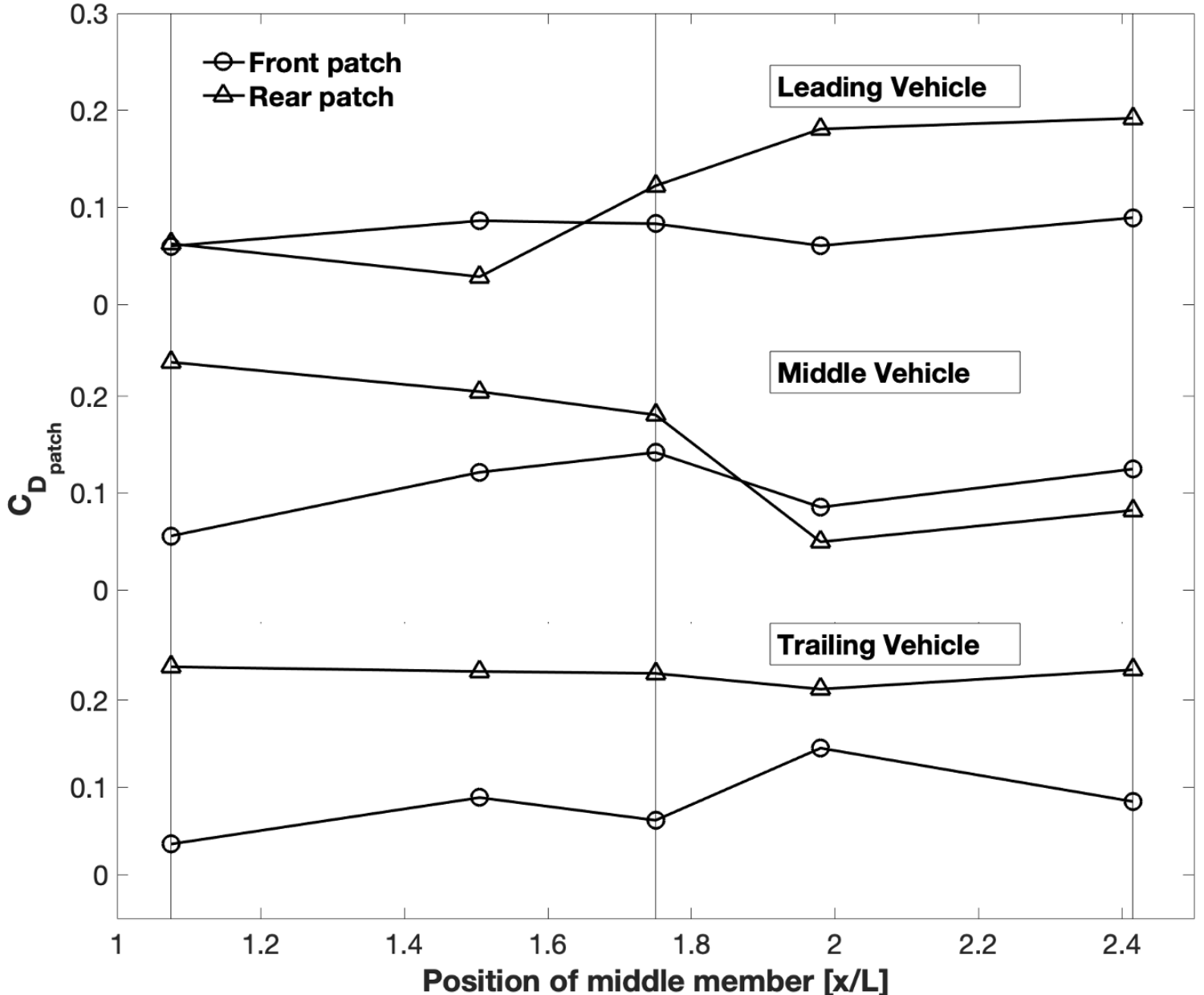}
\caption{Comparison of drag experienced by various patches of vehicles in steady time-averaged platoon cases.}
\label{figure6}
\end{figure}

Figure \ref{figure5} shows the separation distance as the normalized gap between the leading and the middle vehicles based on the vehicle length L and compares the drag of all platoon members with the standalone case. The trailing member does not show significant changes in drag on varying the position of the middle member. However, notable opposite changes are seen between the leading and middle vehicles. Drag decreases across the middle vehicle as it approaches the leading vehicle up to some distance, after which it starts to increase again. This trend is caused by increased pressure difference across it leading to more drag force. When the middle vehicle is placed upstream of the center position, the change is positive, indicating that the vehicle is at a disadvantage in this platoon when compared with the standalone vehicle. This was also observed by Jacuzzi and Granlund \cite{jacuzzi2019} who reported a drag peak for a trailing vehicle with a small gap to the one in front. A reasonable explanation for this can be resonance occurring in such smaller separation gaps \cite{zabat1995}.
Figure \ref{figure5} also shows that the average platoon drag is less than the standalone vehicle for all of the considered positions of the middle vehicle, a relatively constant phenomenon for all of the studied cases. Since the highest combined drag is with the middle vehicle in the centered position, there is no need for active platooning control to spend energy maintaining equidistant spacing, as it actually increases total drag.
To find out the factors governing these drag behaviors, patch drag contributions are taken into account for every platoon member. These patches are defined as frontal surface patch, rear surface patch, top surface patch, bottom surface patch, left surface patch, and right surface patch. For a given patch, drag coefficient is calculated using its Cp distribution and surface area from CFD. At every point on a surface, coefficient of pressure is multiplied with its x-normal and the product is integrated over the surface. Dividing this integration with the respective patch surface area gives the patch drag coefficient as per the following equation.
\begin{equation}
    C_{D,patch}=\frac{C_{p,patch}N_x}{A_{patch}}
\end{equation}
Here, $C_{D,patch}$ is the coefficient of drag for the given patch, $C_{p,patch}$ is the pressure coefficient of that patch, $N_x$ is the surface normal in the X-direction, and $A_{patch}$ is the surface area of the patch. Since the CFD data corresponds to various points on the vehicle surfaces, the product of pressure coefficient $C_p$ and surface normal in the x-direction $N_x$ for every point was integrated over the patch to get a final value. Patch drag coefficient is useful to get better insights into drag generating regions. This difference can be seen in Figure \ref{figure6}, which shows the drag experienced by various patches of a given vehicle.
It should be noted that for all the platoon vehicles, only respective front and rear surfaces contribute to the drag variation. When the middle vehicle approaches the leading vehicle from center position, its front as well as rear surfaces show an increase in the drag. This increment on decreasing the separation distance is backed by the previous studies \cite{zabat1994,zabat1995,jacuzzi2019}. Front surface drag of the middle vehicle increases because of the oncoming flow from the leading vehicle in the closing gap, whereas the rear surface loses pressure as a result of the increasing gap between itself and the front surface of the trailing vehicle. This overall leads to increased pressure difference across it, thus leading to increased aerodynamic drag.
Leading member drag variation shown in Figure \ref{figure5} can be explained by its rear surface drag variation. Notably, the front surface of the leading member does not experience any drag fluctuation as the freestream flow is directly coming onto it. Therefore, its overall drag variation is governed by its rear surface only. This is shown in Figure \ref{figure6}. Moreover, this surface exhibits the opposite drag behavior to that of the middle vehicle front surface.  
Similar to the leading vehicle, the trailing vehicle drag variation is the result of the drag fluctuation on its front surface as it does not have any member behind it. Therefore, as seen in Figure \ref{figure6}, its rear surface experiences almost the same drag in all the cases. Also, the trailing vehicle does not show any significant drag variation when the middle vehicle is placed upstream of its center position. It is only when the middle vehicle is placed downstream of its center position that the trailing member drag varies. This is due to the direct oncoming flow from above and around the middle vehicle on the trailing front surface which leads to increased drag. For the closest gap between the middle and the trailing vehicles, the inwash on the trailing front decreases, therefore, it shows a decrease in the drag.
\begin{figure}
\centering
\includegraphics[width=\columnwidth]{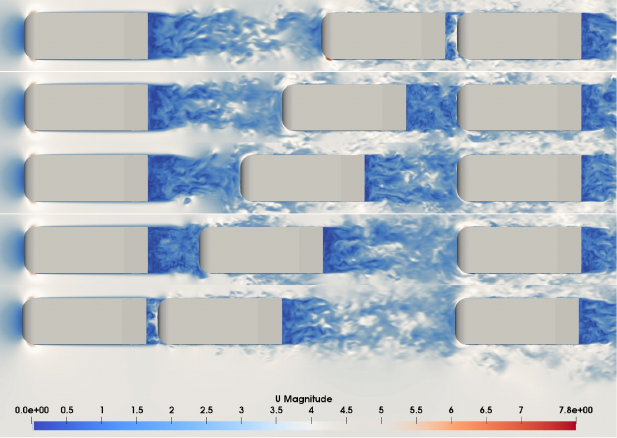}
\caption{Top view of velocity data for steady platoon cases (Cases 1 to 5 from top to bottom, respectively) [complementary video attached]}
\label{figure7}
\end{figure}
Top view of velocity magnitude (m/s) distribution is shown in Figure \ref{figure7} for all the steady platoon cases considered. In all the cases, velocity distribution in the immediate vicinity of the front surface of the leading vehicle and the rear surface of the trailing vehicle is the same. As the middle vehicle approaches the leading vehicle, the flow in their closing gap gets pressurized which is shown in Figure \ref{figure7} by decreasing velocity, i.e. the confined flow is becoming stagnant in the smaller wake. This pressure rise on the leading rear surface and the middle front surface is also shown graphically in Figure \ref{figure6}. As explained earlier, the flow distribution across the trailing member remains almost the same when the middle vehicle is closer to the leading vehicle. In case 2, the stagnant region between the middle and trailing vehicles can be seen in Figure \ref{figure7}.   
Pressure difference across the platoon vehicles accounted for about 88\% of the total drag. The remaining drag is due to the skin friction This also resonates well with Ahmed’s original studies \cite{ahmed1984}, which concluded that the bluff body showed 85\% of the pressure drag and the remaining drag was due to skin friction. Actual cars would experience a higher Reynolds number due to larger size and higher velocity, further diminishing the effect of skin friction contra pressure drag, thus concluding that pressure drag is the dominant effect.
\subsection{Oscillating middle vehicle in a platoon}
\begin{figure}
\centering
\includegraphics[width=\columnwidth]{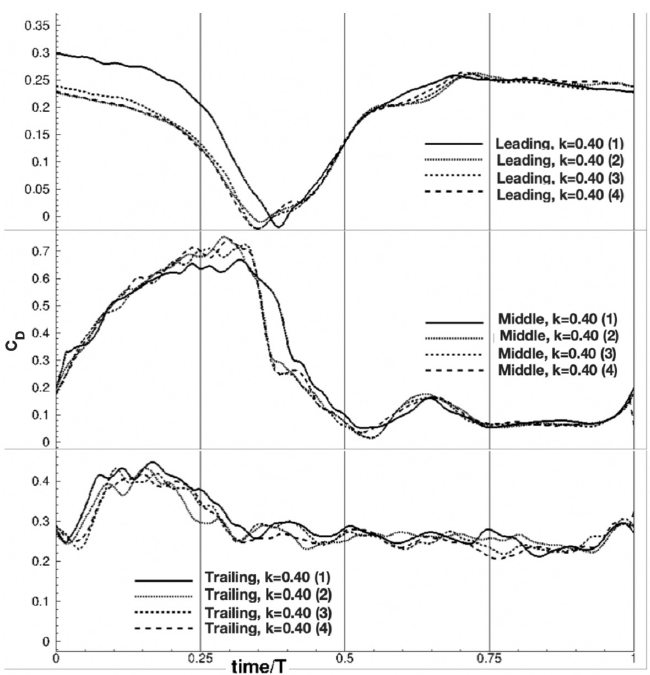}
\caption{Variation of drag coefficient during the four cycles of oscillation of the middle member reduced frequency k=0.4 case.}
\label{figure8}
\end{figure}

\begin{figure}
\centering
\includegraphics[width=\columnwidth]{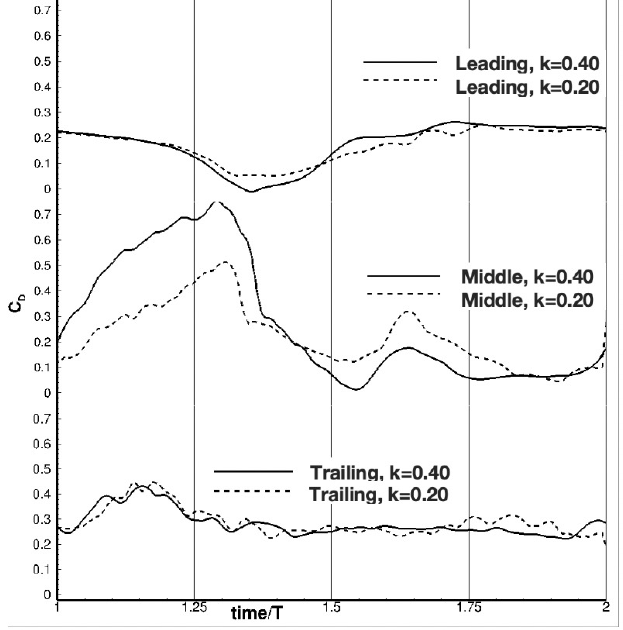}
\caption{Drag coefficient for the leading, middle and trailing vehicles during the second cycle}
\label{figure9}
\end{figure}

\begin{figure}
\centering
\includegraphics[width=\columnwidth]{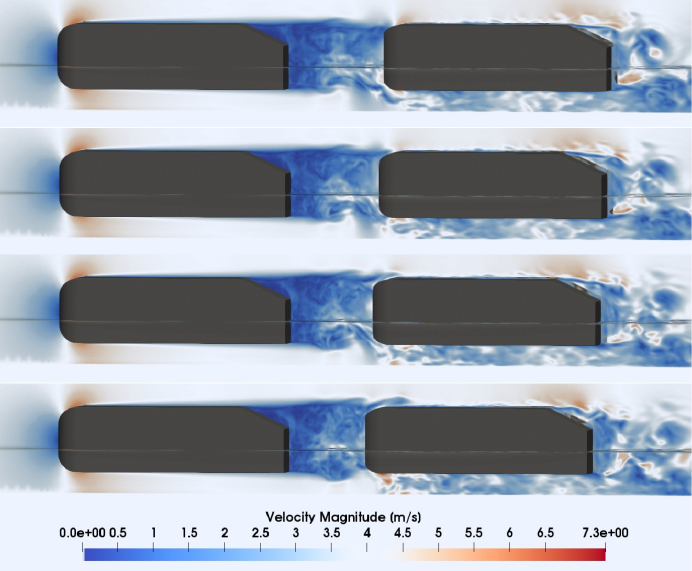}
\caption{Visualization of velocity magnitude in the closing gap as the middle member approaches the leading member during reduced frequency k: 0.4 case. From top to bottom t/T= [1.335; 1.340; 1.351; 1.359] [complementary video attached]}
\label{figure10}
\end{figure}

\begin{figure}
\centering
\includegraphics[width=0.75\columnwidth]{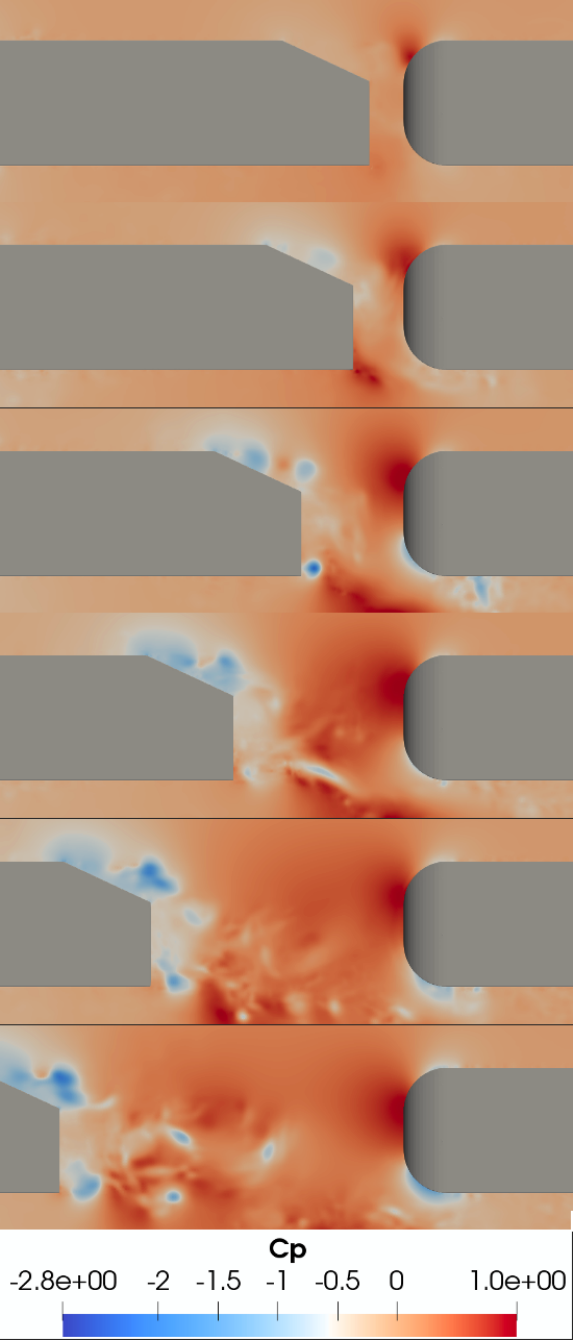}
\caption{Pressure distribution in the increasing gap between the middle and trailing vehicles for reduced frequency k=0.40 case. [From top to bottom t/T= 1.00, 1.05 1.10, 1.15, 1.20, 1.25]}
\label{figure11}
\end{figure}

Simultaneously plotting the four cycles of the overall drag experienced by all the vehicles in the platoon, with middle vehicle oscillating at reduced frequency k=0.4 shown in Figure \ref{figure8}, it is evident that the drag on the trailing vehicle is similar for all cycles in this case. For the leading and middle vehicles, the drag is identical for the last three cycles. Therefore, after one startup cycle the oscillating drag is cyclic and repeatable and the 2nd oscillation results can be used as representative for both oscillatory cases. Similar conclusions about a "startup-time" for oscillatory motion were reached by Granlund et al.\cite{granlund2013b,granlund2014a}
In Figure \ref{figure9}, drag coefficient is normalized with freestream velocity and frontal area and presented for the second cycle for all of the members for two different oscillation frequencies. Time is normalized against the time period of oscillations (T). Oscillating the middle member in a three-member platoon results in oscillating drag for the entire platoon. The second cycle starts with the middle body at the aft most position at t/T=1, highest velocity passing the center at t/T=1.25 and foremost position at t/T=1.5. Drag oscillations of a given member are out of phase with other platoon members at the same reduced frequency. For the leading vehicle, the higher reduced frequency case has a slightly higher amplitude but, other than the peaks, drag variation is almost the same in both cases. For the middle member, increasing the reduced frequency, which is based on the oscillation frequency, increases the drag amplitude, whereas the trailing member does not show any definitive changes in drag. The phase of drag variations remains the same for the respective members in both cases. Therefore, drag peaks are observed at the same location (or at the same time/T instance) for respective members irrespective of the reduced frequencies.
Leading member drag variation presented in Figure \ref{figure9} shows that the minimum drag does not occur when the separation gap is the smallest. This gap is achieved at the normalized time t/T= 1.5 for the second cycle. The leading member drag coefficient reaches a minimum of -0.020 at t/T=1.34 and, on further decreasing the gap, drag coefficient starts to increase. This phase difference in the leading drag and the motion of the middle member is due to the pressure change in the wake of the leading member. As the middle member approaches the leading member, it pressurizes the flow confined in the separation gap and, thus, increases the pressure on the leading rear surface. This ramming effect generates a pressure difference which in effect creates thrust on the leading member in the reduced frequency k=0.4 case. This is plotted as negative drag. Further closing the gap cannot confine the pressurized flow as shown in Figure \ref{figure10}. Therefore, it experiences an increased pulling force, which is drag.
When the middle member accelerates from the initial position, pressure on its rear surface is reduced as the gap confining the wake between itself and the trailing member increases. On the other hand, its front surface pressure increases due to the closing gap between the leading and middle members which is consistent with other published studies \cite{zabat1995,jacuzzi2019}. This increases the pressure differential across it and results in increased drag for the first half-cycle, shown in Figure \ref{figure9}.  Ideally, the maximum pressure should occur at the closest gap as the middle front surface will experience the maximum force. However, it is not the case. Pressure on the middle front surface reaches the maximum and then decreases as the gap is closed further. This pressure loss, shown in Figure \ref{figure10}, is a result of small separation distance which fails to hold the pressurized wake. During this time, the front surface of the middle body experiences negative drag, i.e. a force in the direction of motion because, at this instant, the middle member enters the wake of the leading member. This loss in pressure occurs first on the front surface of the middle member and then on the rear surface of the leading member which can be seen in Figure \ref{figure10}. Therefore, the drag peak on the middle member occurs just before the drag peak on the leading member. As the respective surfaces of the middle and leading members experience pressure loss at slightly different times, when the separation gap between them is closing, a slight shift in the drag peaks is observed for these members. This shift becomes less pronounced for the 0.2 reduced frequency case. However, for this case, the middle member undergoes pressure loss just before the leading member as observed for the higher frequency case. After reaching the extreme front position, as the middle member starts to move backwards, it creates suction in the gap between itself and the leading member. As a result, its front surface is subjected to the flow coming from over the leading member exerting pressure. This increment can be observed in Figure \ref{figure9} at the normalized time of 1.55. Pressure increases for a short period of time after which it decreases. This pressure loss occurs as the middle member leaves the leading wake, where this oncoming flow is absent resulting in reduced drag. Upon approaching the trailing member, wake behind the middle member gets pressurized, increasing its base pressure. Moreover, its front surface also experiences increased pressure. This results in a nearly constant pressure difference and, hence, nearly constant drag. 
Drag variation for the trailing member remains constant after the first quarter cycle in each case. This is due to the trailing member always remaining in the wake of the middle member and an absence of forward acceleration against the flow due to imposed motion. At the end of the cycle, however, when the middle member is very close to the trailing member, both show increased drag. The middle member loses base pressure for very small gaps as explained earlier. However, pressure continues building up on the trailing front surface. This pressure arises because of the direct oncoming flow over and around the middle member as shown in Figure \ref{figure11}.
As the middle vehicle moves forward, pressure in the increasing gap increases as the flow from over the middle vehicle back slant enters the gap. This oncoming flow directly collides on the frontal surface of the trailing member thereby increasing its overall drag.
\begin{figure}
\centering
\includegraphics[width=\columnwidth]{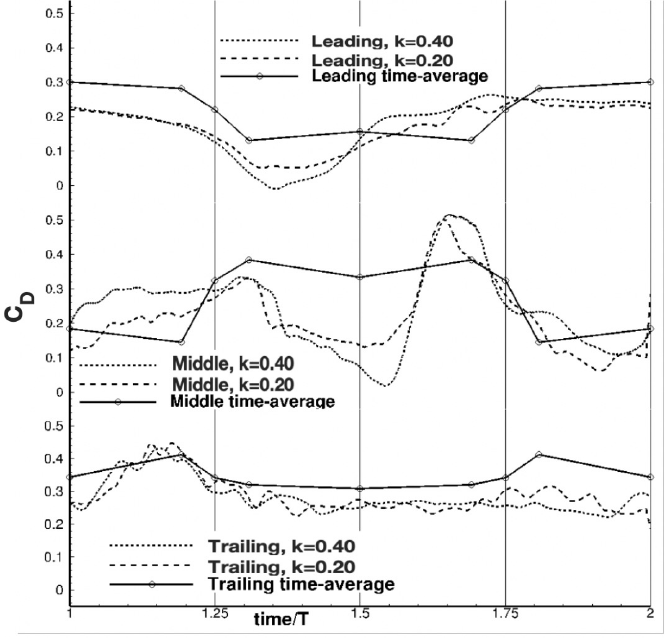}
\caption{Comparison of steady time-averaged drag and the drag normalized with the instantaneous speed for the second cycle.}
\label{figure12}
\end{figure}
To better compare unsteady motions with steady-state results, drag coefficient of the middle vehicle is instead normalized with the instantaneous velocity\cite{granlund2016a}, since it is oscillating, as per the following equation.
\begin{equation}
    C_D=\frac{2 D}{\rho U_{inst}^2 A_f}
\end{equation}
Where D is the drag force, $U_{inst}$ is the instantaneous velocity, and $A_f$ is the frontal area of the Ahmed body. Time-averaged drag for a steady-state vehicle is used to compare with the fluctuating drag for a moving vehicle. Figure \ref{figure12} shows a difference in the steady time-averaged drag and the transient drag. The Lead member experiences lower drag for the first half-cycle than the steady drag. This is because of the pushing effect the middle member has on the flow. It pushes the flow onto the rear surface of the front member and, thus, the pressure difference across it goes down. When the middle vehicle is steady, i.e. not oscillating, this force is absent, which results in higher pressure difference across the leading member. When the middle member translates backwards, it creates a suction force which tries to pull the leading member towards it and results in increased drag. Since the middle member does not oscillate in the steady case, this force does not act on the leading member. Thus, this case shows reduced drag. Similar trends are observed for the middle member as well. As it approaches the leading member, oncoming flow is pressurized, leading to increased drag before losing this pressure as a result of the small gap. Since this movement does not occur in the steady case, drag increases as the middle member is placed close to the leading one. For the second half-cycle, the oscillating member creates suction which leads to increased pressure difference across it. This does not occur in the steady case as there is no member moving backwards. Therefore, we see as much as 75\% difference in the steady and transient results. This is the maximum drag difference and it varies for every member throughout the cycle. Trailing member drag trend looks similar for both steady as well as transient cases. This is because the drag experienced by the trailing member in the transient case is not dependent on dynamic forces, except at the end of the cycle. Instead, it is driven by the position of the middle body and its wake. Both the cases show increased drag in the first quarter-cycle which is due to the flow coming directly onto the trailing member. We see a difference in the plots towards the end as this part is mainly governed by the dynamic forces. Since the middle member tries to pressurize its wake while moving backward, it resists the top flow from entering the wake. Therefore, even though the trailing member comes inside the wake of the middle member, there is no oncoming flow onto its surface. This is the reason transient drag is lower than the steady drag at this location and shown in Figure \ref{figure13}.
\begin{figure}
\centering
\includegraphics[width=0.75\columnwidth]{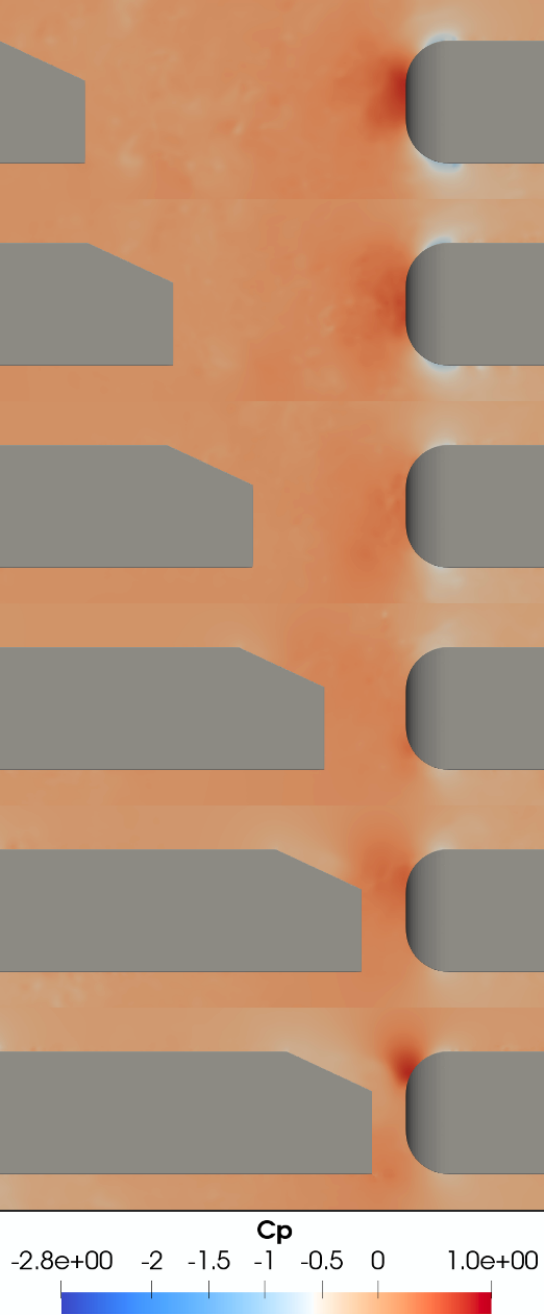}
\caption{Pressure distribution in the decreasing gap between the middle and trailing vehicles for reduced frequency k: 0.40 case. [From top to bottom t/T= 1.75, 1.80 1.85, 1.90, 1.95, 2.00]}
\label{figure13}
\end{figure}
On comparing Figures \ref{figure10} and\ref{figure12}, the difference in the pressure distribution in the gap can be clearly seen. In Figure \ref{figure11}, as the gap increases, a low-pressure region behind the middle vehicle is created and increases with the gap. This low-pressure region sucks the flow inside the gap. On the other hand, when this gap is decreasing, shown in Figure \ref{figure13}, the flow gets slightly compressed, which resists the motion of the flow entering the gap. It is not until the closest gap between the trailing and middle vehicles when the flow enters the gap, shown in the bottom-most case in Figure \ref{figure13}. This is again a dynamic effect, because at this point the middle vehicle has zero velocity. In other words, it stops compressing the flow.

\subsection{Prediction of dynamic drag effects from steady time-averaged data}
The main goal here is to be able to use the static drag data and predict the dynamic effects without running dynamic simulations. Transient forces occurring in such dynamic cases are accounted for and added accordingly in the static cases. These transient forces can be dynamic pressure generated because of the instantaneous velocity of the vehicle and added mass force which is a result of the acceleration of the vehicle and the surrounding fluid. These expressions are specifically applied to the middle vehicle as it is oscillating. Brennen \cite{brennen1982} studied the added mass and inertial fluid forces for various geometries and came up with some empirical formulae to give the approximate added mass of the fluid for a given geometry. Of some of the simple geometry shapes studied are, spheres, cylinders, flat plate, cuboids, etc. As the Ahmed body is a generic bluff body, which resembles a horizontal cuboid with a nearly flat frontal surface, Brennen’s results of simple geometries can be directly applied, and we choose a horizontally oscillating parallelepiped with the same length, width and height as the Ahmed body.
For a given simple geometry, added mass contribution is added to the static drag to get the total force. This total force is then normalized with the instantaneous velocity based on the vehicle position and compared with the transient drag normalized with the instantaneous velocity given as follows.
\begin{equation}
    D_{total}=D_{steady}+D_{added-mass}
\end{equation}
This predicted transient normalized drag is compared with the normalized drag calculated as per the equation in section 4.2. Here, $D_{steady,time-averaged}$ is the steady time-averaged drag force and $D_{added-mass}$ is the added mass force given by Brennen, 
\begin{figure}
\centering
\includegraphics[width=\columnwidth]{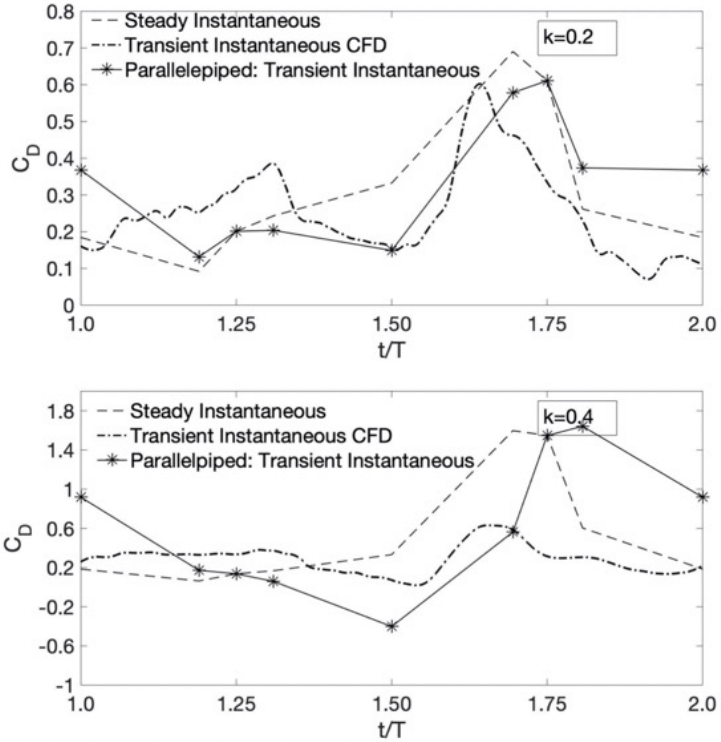}
\caption{Comparison of middle vehicle transient drag predicted using the added mass effects of a parallelepiped with its transient drag from CFD (k= 0.2: top, k= 0.4: bottom)}
\label{figure14}
\end{figure}
Figure \ref{figure14} shows the comparison of the transient drag coefficient calculated from the static drag, which accounts for the added mass effects, with the transient drag coefficient from CFD runs. Added mass effects of a rectangular parallelepiped are co-plotted for the two reduced frequencies considered. For both the reduced frequencies, added mass force is overpredicted at the beginning and end of the oscillation cycle. These overpredictions are amplified for the higher reduced frequency case owing to the increased acceleration of the added mass. This method considers a constant added mass which does not consider the separation gap volume available to contain this theoretical added mass. Since the transient motion is considered through change in acceleration, the middle vehicle has the highest positive acceleration at the start/end point of the cycle. Therefore, more added mass is overpredicted in this case. From Figure \ref{figure14}, when the drag of the oscillating member in a platoon is normalized with its instantaneous velocity and the added mass effects are accounted for, it still does not give an accurate prediction at all the positions during the oscillation. This is due to the convective nature of fluid mechanics, where flow separation requires a finite time to form and subsequently affect the downstream vehicle.
\section{Conclusions}
This study concluded that the transient CFD tests do not give the same results as that of the steady CFD tests. Steady platoon with the middle vehicle placed at the center experienced the highest average drag, therefore maintaining equal separation distance in a platoon of three vehicles may not always give minimum overall platoon drag. However, the overall platoon drag for all the steady cases considered here was less than that of the steady-standalone vehicle at the same Reynolds number. When a platoon vehicle is under oscillations, it produces dynamic forces such as suction and compression. A steady analysis, where the middle vehicle is simply placed at different static positions mimicking the dynamic motion, fails to account for these dynamic forces leading to as much as 75\% difference in the drag. This magnitude varies for each member and depends on the position of the oscillating member. The drag behavior of a platoon member is governed by its surfaces which have a vehicle present in their vicinity and are perpendicular to the freestream. The leading member experienced nearly the same drag for both the frequencies with the exception of the smallest gap. At this point, higher reduced frequency resulted in a higher ramming effect which increased the base pressure of the lead member and decreased its drag. For the middle member, increasing oscillations resulted in increased drag amplitude with the same phase. Pressure loss phenomenon was observed when the gap between the lead and the middle member was too small to confine the pressurized wake. Considering the added mass effects of a flat plate or cylinder in the steady CFD results does not accurately predict the transient results at every position. This variation comes from the convective nature of the fluid which is not considered in added mass corrections. Therefore, the effect of oscillations in the platoon should be studied with a transient approach to account for the real-time dynamic effects. High speed racing with very small separation gaps, also known as drafting, is observed in NASCAR stock cars and other racing forms. Since these race cars are manually operated, they are very likely to undergo oscillations. This study allows for the greater understanding of the dynamic effects of drag.

\bibliographystyle{plain}
\bibliography{references.bib}

\end{document}